\documentclass[review,12pt]{elsarticle}

\usepackage{amssymb}
\usepackage{amsmath}

\usepackage{orcidlink}
\newcommand{\bs}[1]{\boldsymbol{#1}}

\journal{Acta Astronautica}

\begin{document}

\begin{frontmatter}

\title{Short-Term Evolution and Risks of Debris Cloud Stemming from Collisions in Geostationary Orbit}

\author[1]{Peng Shu \orcidlink{0000-0002-5970-9992}}
\author[2]{Meng Zhao \orcidlink{0009-0000-2231-9715}}
\author[3]{Zhen-Yu Li \orcidlink{0000-0001-7249-962X}}
\author[3]{Wei Sun}
\author[1]{Yu-Qiang Li}
\author[2]{Ya-Zhong Luo \corref{cor1}}
\ead{luoyz@nudt.edu.cn}
\cortext[cor1]{Corresponding author}

\affiliation[1]{organization={Yunnan Observatories, Chinese Academy of Sciences},
postcode={650216},
city={Kunming},
country={China}}
\affiliation[2]{organization={National University of Defense Technology},
postcode={410073},
city={Changsha},
country={China}}
\affiliation[3]{organization={Beijing Institute of Tracking and Telecommunications Technology},
postcode={100094},
city={Beijing},
country={China}}

\begin{abstract}
  The increasing population of objects in geostationary orbit has raised concerns about the potential risks posed by debris clouds resulting from fragmentation. 
  The short-term evolution and associated hazards of debris generated by collisions in the geostationary region is investigated in this study.
  The initial distribution of two debris clouds is modeled using a single probability density function. 
  The combined distribution of the evolved clouds is determined by solving boundary value problems. 
  The risks associated with these debris clouds are evaluated by calculating the instantaneous impact rate and cumulative collision probability. 
  The probability of collisions with millimeter-sized fragments may increase to 1\% within 36 hours, while the probability of collisions with fragments 5 cm or larger is approximately $10^{-5}$. 
  These findings underscore the vulnerability of the geostationary region to space traffic accidents.
\end{abstract}


\end{frontmatter}



\section{Introduction}\label{sect:intro}

The geostationary orbit is a popular orbit for communication, meteorological, and navigation satellites due to its apparent motionless.
Nearly all geostationary satellites are positioned in a circular orbit with a radius of  42,164 km, making this region particularly vulnerable to space traffic accidents due to the high concentration of objects \cite{esasspacedebrisoffice2018Classification} and the absence of natural debris-clearing mechanisms \cite{mei2022Hybrid}. 
The growing population in geostationary region raises concerns about the potential risks posed by fragments stemming from explosions and collisions, particularly following the breakup of Intelsat-33e, which remained operational in geostationary orbit  until October 19, 2024 \cite{jewett2024Intelsats}.

A breakup event generates a large number of fragments of varying sizes \cite{anz-meador2022History}. In the geostationary region, only fragments larger than 1 meter are routinely tracked by the Space Surveillance Network \cite{nafi2020Practical}, as the sensitivity of ground-based sensors decreases significantly with distance \cite{blake2021DebrisWatch}. However, small, non-trackable fragments can still cause catastrophic damage to spacecraft \cite{mcknight2019Space}. The collision velocity of spacecraft in geostationary orbit can reach up to 4 km/s \cite{oltrogge2018Comprehensive}, while micro-meteoroids may hit at speeds of up to 72 km/s \cite{programs2011Limiting}.

The impact of a debris cloud is inherently global \cite{lawrence2022Case} as it disperses around the entire Earth. 
Several engineering tools have been reported by the Interagency Space Debris Coordination Committee for the risk assessment of space debris \cite{2018IADC}.
Among them, NASA’s Orbital Debris Engineering Model \cite{matney2019NASA,krisko2014New} and ESA’s Meteoroid and Space Debris Terrestrial Environment Reference (MASTER) \cite{horstmann2020Enhancement} are considered state-of-the-art models for their respective agencies \cite{krisko2015ORDEM}.
Harbin Institute of Technology has also established a Space Debris Environment Engineering Model, which yield results comparable to MASTER-8  \cite{liu2024Space}.
Most of these engineering tools treat fragments as discrete particles, with their initial states drawn from statistical distributions such as the NASA Standard Breakup Model \cite{krisko2011Proper}, the IMPACT fragmentation model from The Aerospace Corporation\cite{mains2022IMPACT}, and the Spacecraft Breakup Model from China Aerodynamics Research and Development Center \cite{yao2024Generation}.
To achieve reliable results, large sample sizes and repeated simulations are necessary\cite{jang2024Simulating,liou2006Risks}, as discrete methods lack the sensitivity to capture low-probability events\cite{au2001Estimation}. Consequently, the computational burden is significantly heavy.

Instead of propagating discrete particles, the continuous method propagates the probability density function of debris clouds.
McInnes modeled the debris clouds using the fluid continuity equation \cite{mcinnes1993Analytical}.
Letizia et al. developed an analytical model to evaluate the collison risk of a single fragmentation cloud \cite{letizia2015Analytical,letizia2016Collision}.
Subsequently, Frey and Colombo presented a fully probabilistic framework with the continuous initial distribution derived from breakup models \cite{frey2021Transformation}.
Giudici et al. introduced density-based methods into a complex evolutionary model to predict the space environment over 200 years \cite{giudici2024Densitybaseda}.
Recently, Wen et al. developed an inverse mapping method that accounts for J2 perturbation, enabling the evaluation of mid-term evolutions \cite{wen2024Modeling}.
However, simplifications in distributions and orbits are required in these models, making them less competitive for the short-term evolution which is fast-changing and inhomogeneous.

In recent years, the idea of evaluating debris cloud with boundary value problems (BVPs) has been revived to capture the accurate distribution of short-term clouds.
Healy et al. developed a BVP-based method to demonstrate the exact spatial density of tens of orbits \cite{healy2016Structure,healy2020Orbital}.
Vallado and Oltrogge introduced the Debris  Risk Evolution and Dispersal tool of the Analytical Graphics, Inc.\footnote{Its current name is Ansys Government Initiatives.}, which evaluates the three-dimensional distribution of an evolving debris cloud by solving Lambert problems \cite{vallado2017Fragmentation,oltrogge2017Application}.
Shu et al. atiopresented the joint distribution of positons and velocities of a fragmentation cloud, and derived the impact risk from this distribution \cite{shu2022Collision,shu2023Impact}.
Parigini et al. applied the automatic domain splitting technique to reduce the computational cost of calculating collison probability over time \cite{parigini2024ShortTerm}.

In this work, the BVP-based methods was enhanced by representing two clouds as a single probability density function. 
The initial density function was derived from NASA Standard Breakup Model as continuous functions, and the summed distribution was propagated by solving boundary value problems.
The remainder of this paper is organized as follows. 
Section \ref{sec:initial_distribution} derives the initial distribution of two debris clouds. 
The evolution of the debris cloud is described in Section \ref{sec:evolution}. 
The hazards posed by the two debris clouds in the geostationary region are evaluated in Section \ref{sec:risks}. 
Finally, conclusions are drawn in Section \ref{sec:conclusions}.

\section{Problem Statement}

\begin{figure}[htbp]
  \centering
  \includegraphics[width=0.5\textwidth]{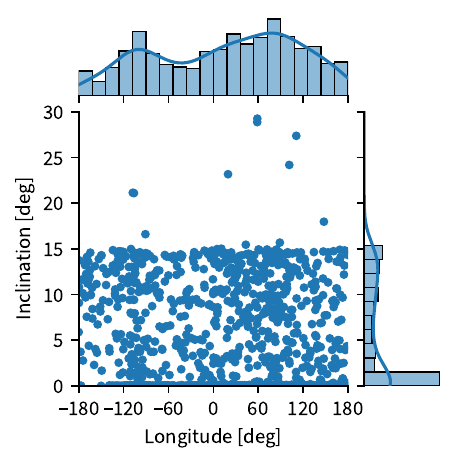}
  \caption{The distribution of the longitude and inclination of GEO objects.}
  \label{fig:gsodist}
\end{figure}

By 2024, over 1,000 objects have been observed near the geostationary orbit (GEO) \footnote{Orbit elements downloaded from www.space-track.org with mean motion between [0.99,1.01] and eccentricity less than 0.01.}.
The longitude and inclination of these objects are shown in Figure \ref{fig:gsodist}. 
Nearly all objects exhibit inclinations of less than 15 degrees, with the majority having inclinations of less than 1 degree. 
Once a fragmentation event occurs, the GEO objects will be exposed to considerable risks,  as they are densely clustered along a single ring above the Equator.

\begin{figure}[htbp]
  \centering
  \includegraphics[width=0.5\textwidth]{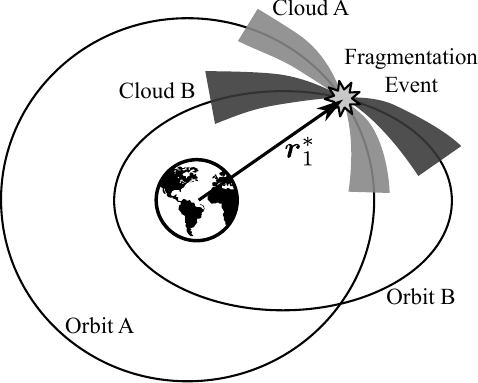}
  \caption{Illustration of fragmentation events.}
  \label{fig:collisionfrag}
\end{figure}

Consider a collison of two objects at position $\bs{r}_1^*$ at epoch $t_1$, esulting in an instantaneous fragmentation event that generates two debris clouds, as shown in Figure \ref{fig:collisionfrag}. 
Our objective is to determine how these clouds spread and the associated risk to other GEO objects.

\section{Initial Distributions of Fragmentation Clouds} \label{sec:initial_distribution}

\subsection{The Initial PDF of A Fragment}
Consider a randomly selected fragment within the clouds, with its initial position denoted as $\bs{r}_1$. 
Since all fragments are ejected from a single point, the probability density function (PDF) of $\bs{r}_1$ can be represented using the Dirac delta function \cite{shu2023Impact}:
\begin{equation}
  p_{\bs{r}1}(\bs{r}_1) = \delta (\bs{r}_1 - \bs{r}_1^*)
\end{equation}
where $\delta(x)$ is the Dirac delta function satisfying
\begin{equation}
    \label{eq:dirac_func_def}
    \begin{gathered}
        \int_{-\infty}^{+\infty} \delta(x) \, \mathrm{d} x  = 1                         \\
        \delta(x)                                           = 0 \quad \text{if }  x \neq 0
    \end{gathered}
\end{equation}

The initial velocity of the fragment can be obtained by adding the ejection velocity to that of its parent object,
\begin{equation}
  \bs{v}_1  = \bs{v}_0 + \Delta \bs{v}
\end{equation}
where $\bs{v}_0$ represents the parent object's velocity, and $\Delta \bs{v}$ denotes the ejection velocity. The probability density function of the initial velocity can then be expressed as
\begin{equation}
  \label{eq:pv1_sum}
  p_{\bs{v}1} (\bs{v}_1) = \frac{N_a}{N_a + N_b}
    p_{\Delta \bs{v}a}(\bs{v}_1 - \bs{v}_{a0}) + \frac{N_b}{ N_a + N_b} p_{\Delta \bs{v}b}(\bs{v}_1 - \bs{v}_{b0})
\end{equation}
where $N_a$ and $N_b$ are the number of fragments in clouds A and B, respectively. $p_{\Delta \bs{v}a}$ and $p_{\Delta \bs{v}b}$ are the PDF of the ejection velocity of two clouds, which can be derived from breakup models (Sec.\ref{sec:pdf_ejection}). 

Thus, the PDF of the initial state can be obtained by combining the PDFs of the position and velocity:
\begin{equation}
  \label{eq:prv1}
  p_{\bs{r}_1,\bs{v}_1} (\bs{r}_1, \bs{v}_1) = \delta (\bs{r}_1 - \bs{r}_1^*) p_{\bs{v}1} (\bs{v}_1)
\end{equation}

\subsection{The Distribution of Ejection Velocity} \label{sec:pdf_ejection}

The NASA Standard Breakup Model (SBM) is a widely used model for predicting the initial distribution of fragmentation clouds \cite{jang2024Simulating,letizia2015Analytical,frey2021Transformation,giudici2024Densitybaseda,healy2016Structure}.
It comprises a set of simple empirical equations that describe the size, area-to-mass ratio, and ejection velocity of the generated fragments, yielding highly reliable statistical results in catastrophic fragmentation scenarios \cite{francesconi2019CST}. 
This simplification ignores the effects of factors such as impact direction, material properties, and structural characteristics \cite{olivieri2024Numerical}.
Recently, the University of Padova developed the Collision Simulation Tool (CST) to numerically evaluate in-space fragmentation events\cite{francesconi2019CST}. 
This tool models colliding objects with a mesh of macroscopic elements, allowing for the simulation of complex impact configurations. 
Nonetheless, the focus of this paper is to assess the consequences of general catastrophic collisions in geostationary orbits, rather than being specific to particular spacecraft structures and materials. Consequently, the NASA SBM is employed in the subsequent analysis.

\subsubsection{Conditional Probability Functions of NASA SBM}

In typical applications, breakup models are employed to generate a large number of discrete sample fragments, with initial states following specific statistical distributions. 
In contrast, our approach seeks a continuous distribution function that facilitates analytical transformation.

In NASA SBM, the number of fragments with characteristic length larger than $L_c$ is
\begin{equation}
    \label{eq:number_length}
    N(L > L_c) = k L_c^{-\beta}
\end{equation}
where $k$ and $\beta$ are unitless parameters determinted by the type  of fragmentation and parent objects. 
The PDF of $L$ in the interval $[L_{\min},L_{\max}]$ can then be derived as \cite{frey2021Transformation}:
\begin{equation}
    \label{eq:pdf_length}
    p_L(L) = \beta \frac{L^{-(\beta+1)}}{L_{\min}^{-\beta}-L_{\max}^{-\beta}}
\end{equation}
where $L_{\max}$ and $L_{\min}$ represent the upper and lower bounds of the characteristic lengths, respectively.
The distribution of further parameters in NASA SBM are usually expressed with the logarithm of characteristic length $\lambda = \log_{10}(L)$, which PDF can be written as
\begin{equation}
  \label{eq:pdf_lmd}
  p_{\lambda} (\lambda) = \ln(10)10^{\lambda} p_L(10^{\lambda})
\end{equation}

The area-to-mass ratio $A/m$ of a fragment is determinted by a conditional PDF:
\begin{equation}
  \label{eq:pdf_chilmd}
  p_{\chi|\lambda} (\chi, \lambda)= \alpha(\lambda) \mathcal{N}\left(\mu_1(\lambda),\sigma_1^2(\lambda)\right) + (1-\alpha(\lambda)) \mathcal{N}\left(\mu_2(\lambda),\sigma_2^2(\lambda)\right)
\end{equation}
where $\chi = \log_{10}(A/m)$, $\mathcal{N}$ is the PDF of the normal distribution, and the functions $\mu_1$, $\sigma_1$, $\mu_2$, $\sigma_2$ and $\alpha$ are determined by the type of parent object.

The ejection velocity $\Delta v$ is also determined by a conditional PDF in NASA SBM:
\begin{equation}
  \label{eq:pdf_uchi}
  p_{u|\chi} (u, \chi)= \mathcal{N}\left(\mu_u(\chi),\sigma_u^2(\chi)\right)
\end{equation}
where $u = \log_{10}(\Delta v)$ and the functions $\mu_u$ and $\sigma_u$ are determined by the type of fragmentation. 

\subsubsection{The Distribution of Directional Ejection Velocity}

The marginal distribution of $u$ can then be calculated by integrating Eqs. \eqref{eq:pdf_lmd}-\eqref{eq:pdf_uchi}:
\begin{equation}
  \label{eq:pu}
  p_u(u) = \int_{\chi_{\min}}^{\chi_{\max}} p_{u|\chi} (u, \chi) \int_{\lambda_{\min}}^{\lambda_{\max}} p_{\chi|\lambda} (\chi, \lambda) p_{\lambda} (\lambda) \;\mathrm{d} \lambda \;\mathrm{d} \chi
\end{equation}
where $\lambda_{\min}$, $\lambda_{\max}$, $\chi_{\min}$ and $\chi_{\max}$ are determined based on the boundaries of the characteristic lengths and the area-to-mass ratio.

Then the PDF of the ejection velocity can be derived as
\begin{equation}
  \label{eq:pdv}
  p_{\Delta v} (\Delta v) = \frac{p_u\left(\log_{10}\Delta v\right)}{\ln(10)\Delta v}
\end{equation}

The NASA SBM assumes that the direction of the ejection velocity is uniformly distributed in space, so the PDF of the directional ejection velocity is
\begin{equation}
  \label{eq:pdvec}
  p_{\Delta \bs{v}} (\Delta \bs{v}) = \frac{p_{\Delta v} (\Delta v)}{4\pi(\Delta v)^2}
\end{equation}
where $\Delta v = \|\Delta \bs{v}\|$.

\subsection{Numerical Results} \label{sec:num_initial}

Jupiter 3 (also known as EchoStar 24), with a mass of 9,200 kg, is currently the heaviest satellite in GEO. 
Its launch requires a Falcon Heavy rocket. 
We consider a hypothetical collision between the upper stage of the Falcon rocket and EchoStar. The parameters of both objects prior to the collision are provided in Table \ref{tab:parameters}, where "RB" refers to the rocket body and "SC" refers to the spacecraft. 

For simplicity, we define the length unit as $\mathrm{LU} = 42,164$ km which corresponds to the radius of the breakup point, and the time unit as $\mathrm{TU} = \sqrt{\mathrm{LU}^3/\mu}$, where $\mu$ is Earth's gravitational parameter. Consequently, the velocity unit becomes $\mathrm{VU} = \mathrm{LU}/\mathrm{TU}$, and the fragmentation point is located at $\bs{r}_1^* = (1,0,0)$ LU. Additionally, the longitude of the spacecraft’s sub-satellite point at the time of fragmentation is set to 0 degrees.

\begin{table}[htbp]
  \centering
  \caption{Parameters of the collision objects.}
  \label{tab:parameters}
  \begin{tabular}{lcc}
    \hline
    Object & SC & RB \\
    \hline
    Mass (kg) & 9,200 & 20,000 \\
    Semi-major axis (km) & 42,164 & 11,530 \\
    Eccentricity & 0.0 & 0.728 \\
    Inclination (deg) & 0 & 28.5 \\
    Right ascension of ascending node (deg) & 180 & 0 \\
    Argument of latitude (deg) & 0 & 180 \\
    True anomaly (deg) & 180 & 180 \\
    \hline
  \end{tabular}
\end{table}

In this scenario, the relative kinetic energy of the collision is 778 J/g, exceeding the 40 J/g threshold for catastrophic fragmentation \cite{krisko2011Proper}. 
A substantial number of fragment samples are then generated using a validated implementation\footnote{https://github.com/esa/NASA-breakup-model-cpp} of the NASA SBM \cite{schuhmacher2021Efficient}. 
These discrete samples are subsequently compared with the continuous probability density function (PDF). 
The number of fragments in each size category is detailed in Table \ref{tab:size_number}.

\begin{table}[htbp]
  \centering
  \caption{The number of fragments by size category.}
  \label{tab:size_number}
  \begin{tabular}{ccc}
    \hline
    $L$, m  & $N_a$ (SC) & $N_b$ (RB)\\
    \hline
    0.05 - 1.00 & $1.174\times 10^4$ & $ 2.552\times 10^4 $\\
    0.01 - 0.05 & $1.733\times 10^5$ & $ 3.768\times 10^5 $\\
    0.001 - 0.05 & $9.309\times 10^6$ & $ 2.024\times 10^7 $\\
   \hline
  \end{tabular}
\end{table}

\begin{figure}[htbp]
  \centering
  \includegraphics[width=1.0\textwidth]{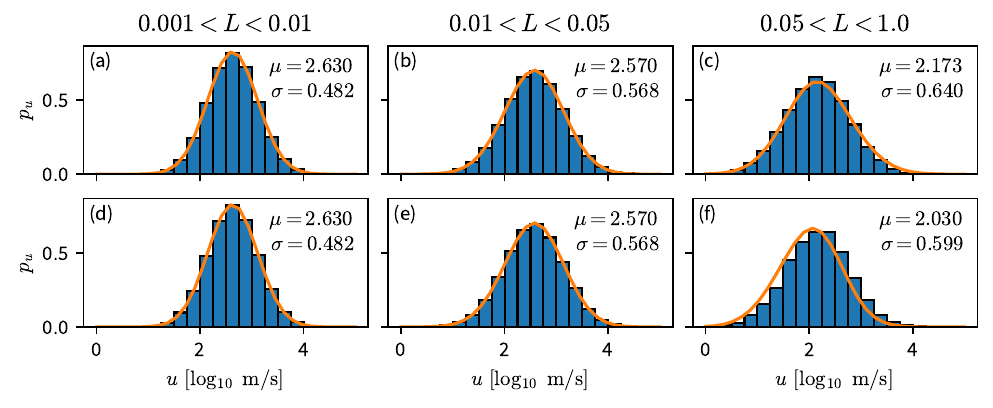}
  \caption{The distribution of ejection velocities by size category.}
  \label{fig:pu}
\end{figure}

Figure \ref{fig:pu} presents the PDF of ejection velocity categorized by size. 
Panels (a)-(c) illustrate the fragment distribution related to RB, whereas panels (d)-(f) depict fragments associated with SC. 
The histogram represents the statistical distribution of discrete samples, and the solid line represents the continuous PDF as defined by Eq.~\eqref{eq:pu}.
The figure demonstrates a strong agreement between the two methods. In each panel, $\mu$ and $\sigma$ denote the mean and standard deviation, respectively, derived from the fragment samples. 
The ejection velocity distribution varies substantially across size ranges, with a marked difference in larger size intervals, where fewer fragments are generated and the distributions from different parent types are notably distinct.

\begin{figure}[htbp]
  \centering
  \includegraphics[width=1.0\textwidth]{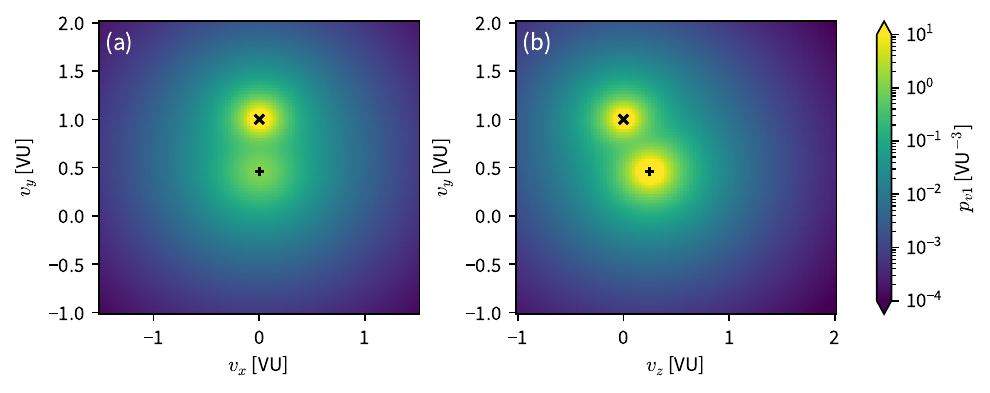}
  \caption{The summed PDF of the directional initial velocity.}
  \label{fig:pdv}
\end{figure}

The distribution of directional ejection velocities for each object is calculated using Eq. \eqref{eq:pdvec}. 
Subsequently, the summed PDF of the two clouds is determined according to Eq. \eqref{eq:pv1_sum}, as illustrated in Figure \ref{fig:pdv}.
The symbol "$\times$" denotes the velocity of SC prior to fragmentation, while "$+$" represents the velocity of RB. 
The probability density is notably higher in the region near the velocity of the parent bodies.
The summed PDF is continuous and smooth, allowing it to be directly employed to form the initial state distribution of any number of fragments,  as outlined in Eq.~\eqref{eq:prv1}.

\section{Evolution of the Cloud Distribution} \label{sec:evolution}

\subsection{The PDF of Fragment State}

After fragmentation, the fragments spread out in new orbits. After a time $t$, the position and velocity of a fragment become
\begin{equation}
  \begin{aligned}
      \bs{r}_2 & = \varphi_r (t, \bs{r}_1, \bs{v}_1) \\
      \bs{v}_2 & = \varphi_v (t, \bs{r}_1, \bs{v}_1)
  \end{aligned}
\end{equation}
where $\varphi_r$ and $\varphi_v$ are related to the orbital dynamics.

The PDFs of $\bs{r}_2$ and $\bs{v}_2$ can then be derived  by the transformation of variables:
\begin{equation}
  p_{\bs{r}_2,\bs{v}_2}(\bs{r}_2,\bs{v}_2) = p_{\bs{r}_1,\bs{v}_1}\left(\varphi_r^{-1}(t,\bs{r}_2,\bs{v}_2),\varphi_v^{-1}(t,\bs{r}_2,\bs{v}_2)\right)  \left|\det\left(\frac{\partial(\bs{r}_2,\bs{v}_2)}{\partial(\bs{r}_1,\bs{v}_1)}\right)\right|^{-1}
\end{equation}
By exploiting the properties of $\delta$ function, this equation can be further simplified to \cite{shu2023Impact}:
\begin{equation}
  \label{eq:prv}
  p_{\bs{r}_2,\bs{v}_2}(\bs{r}_2,\bs{v}_2) = \sum_{i = 1}^{m} \delta(\bs{v}_2 - \bs{v}^*_{2i})  p_{\bs{v}1}\left(\varphi_v^{-1}(t,\bs{r}_2,\bs{v}_2)\right)   \left|\det\left(\frac{\partial \bs{r}_2}{\partial \bs{v}_1}\right)\right|^{-1}
\end{equation}
where
\begin{equation}
  \label{eq:v2i}
  \bs{v}^*_{2i} \in \left\{\bs{v}_2 \mid \varphi_r^{-1}(t, \bs{r}_2,\bs{v}_2) - \bs{r}_1^* = 0\right\}
\end{equation}
involves solving a boundary value problem.

\subsection{The Marginal Distribution of Fragment Position}
Then, the marginal distribution of fragment positions can be directly integrated as:
\begin{align}
  \label{eq:pr2}
  p_{\bs{r}2} (\bs{r}_2) & =  \int_{\mathbb{R}^d } p_{\bs{r}_2,\bs{v}_2} (\bs{r}_2,\bs{v}_2) \,\mathrm{d}\bs{v}_2 \nonumber                                                                                                                                                    \\
                         & = \sum_{i = 1}^{m} \int_{\mathbb{R}^d }\delta(\bs{v}_2 - \bs{v}^*_{2i})  p_{\bs{v}1}\left(\varphi_v^{-1}( t,\bs{r}_2,\bs{v}_2)\right)   \left|\det\left(\frac{\partial \bs{r}_2}{\partial \bs{v}_1}\right)\right|^{-1} \,\mathrm{d}\bs{v}_2 \nonumber \\
                         & = \sum_{i = 1}^{m} p_{\bs{v}1}\left(\varphi_v^{-1}( t,\bs{r}_2,\bs{v}^*_{2i})\right)   \left|\det\left(\frac{\partial \bs{r}_2}{\partial \bs{v}_1}\right)\right|^{-1}
\end{align}
This equation can be written as 
\begin{equation}
  p_{\bs{r}2} (\bs{r}_2) =\sum_{i = 1}^{m} p_{\bs{r}2,i} (\bs{r}_2)
\end{equation}
where
\begin{equation}
  p_{\bs{r}2,i} (\bs{r}_2) = p_{\bs{v}1}\left(\varphi_v^{-1}( \bs{r}_2,\bs{v}^*_{2i})\right)   \left|\det\left(\frac{\partial \bs{r}_2}{\partial \bs{v}_1}\right)\right|^{-1}
\end{equation}
is the density attributed to the $i$-th solution of Eq. \eqref{eq:v2i}.

Finally, we can calculate the density of fragments at position $\bs{r}_2$ at time $t$:
\begin{equation}
  \label{eq:nr2}
  n_{\bs{r}2} = N_f * p_{\bs{r}2} (\bs{r}_2)
\end{equation}
where $N_f = N_a +N_b$ is the total number of fragments.

\subsection{Numerical Results}

The initial distribution obtained in Sec. \ref{sec:num_initial} is transferred using Eq.~\eqref{eq:prv}, and the resulting density distribution after the fragmentation of the two clusters is obtained using Eq.~\eqref{eq:pr2}.    

\begin{figure}[htbp]
  \centering
  \includegraphics[width=0.8\textwidth]{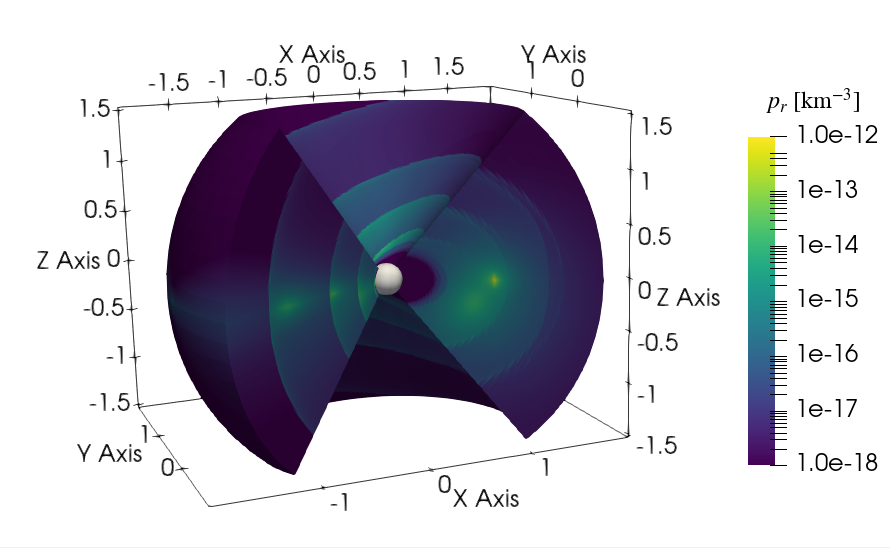}
  \caption{The 3D view of probability density.}
  \label{fig:pr3d}
\end{figure}

Figure \ref{fig:pr3d} depicts the probability density of the two clouds in three-dimensional space 48 hours after the fragmentation.
In this illustration, the white ball at the center represents the Earth.
The color map indicates the probability density of fragments, with yellow signifying higher density and blue indicating lower density. 
This image reveals the layered structure of the fragmentation cloud, which is explored in more detail in subsequent figures.

\begin{figure}[htbp]
  \centering
  \includegraphics[width=1.0\textwidth]{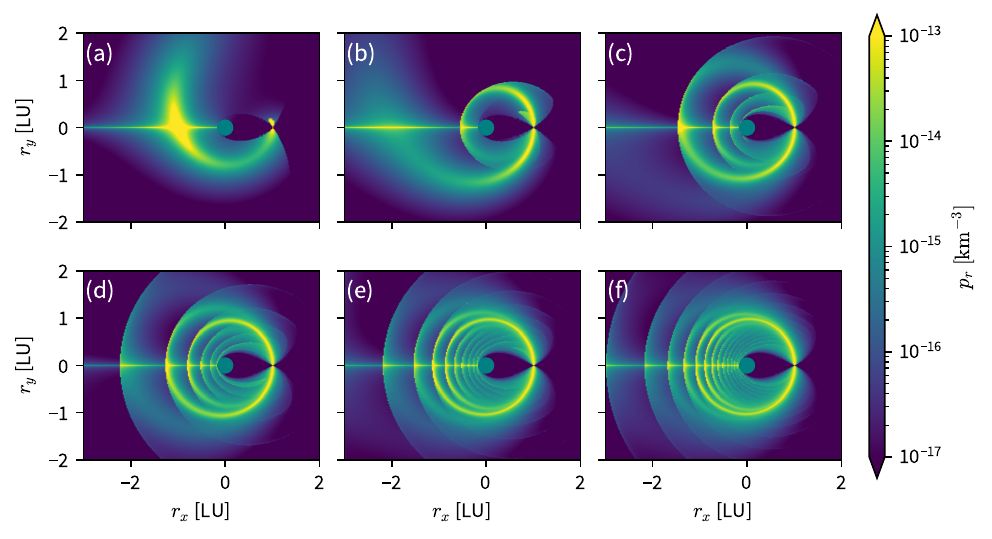}
  \caption{The evolution of density in XY plane.}
  \label{fig:prxy}
\end{figure}

\begin{figure}[htbp]
  \centering
  \includegraphics[width=1.0\textwidth]{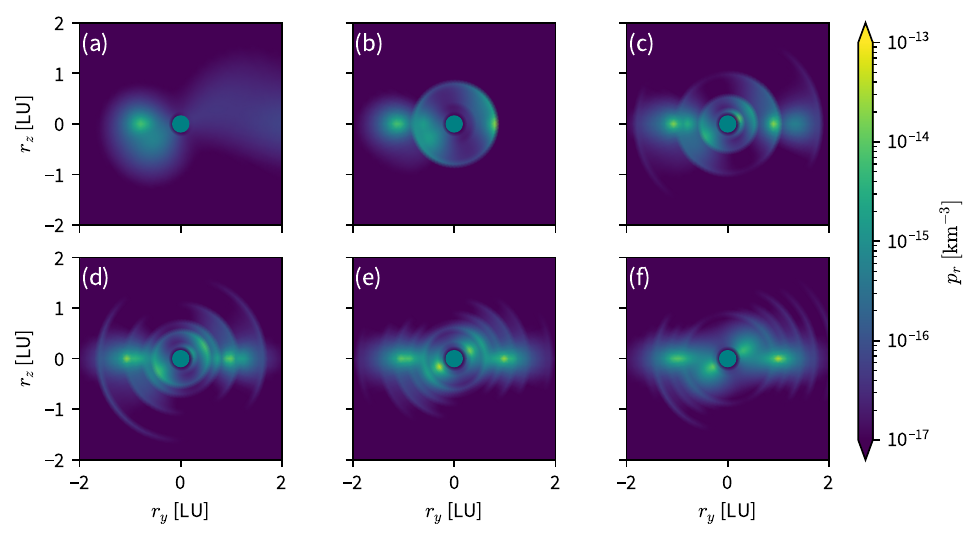}
  \caption{The evolution of density in YZ plane.}
  \label{fig:pryz}
\end{figure}

Figures \ref{fig:prxy} and \ref{fig:pryz} present the probability density in the XY and YZ planes, respectively. 
The green circle at the origin represents Earth, consistent with Fig.\ref{fig:pr3d}. 
Panels (a)-(f) correspond to 0.5, 1, 2, 3, 5, and 7 days after fragmentation.

Figures \ref{fig:prxy} clearly shows a pinch point on the right side of the Earth, exactly where the collision of the two objects occurs.
After 0.5 days, the majority of the fragments have dispersed to the opposite side of the fragmentation point, while a small number have returned to the original location.
After 7 days,  the fragments have spread out to a larger area, forming several complete rings around Earth.
A high-density linear region was observed on the opposite side of the fragmentation point. This is due to the fact that the orbital plane of any fragment will necessarily contain the initial position vector $\bs{r}_1^*$.

Figure \ref{fig:pryz} illustrates the density distribution in the YZ plane, revealing the inclination patterns of the fragments.
Over time, the debris cloud has separated into two distinct clusters in terms of inclination. 
One cluster, located in the equatorial plane, primarily consists of fragments from SC. 
The other cluster, near an inclination of 28 degrees, is mainly composed of fragments from RB.
Moreover, the multi-layered structure of the debris cloud is also clearly visible in the YZ plane.

\section{Risks of the Evolving Debris Clouds} \label{sec:risks}

The evolving debris cloud has the potential to collide with spacecraft in neighboring orbits, posing significant risks, including potentially fatal consequences.
To assess the collision risks, we need to calculate the probability of collision between the debris cloud and a spacecraft.

\subsection{Cumulative Collsion Probability}
Assuming that the position and velocity of a spacecraft are $\bs{r}_s$ and $\bs{v}_s$, respectively, the impact flux it suffers can be calculated from the debris density and velocity \cite{shu2023Impact}:
\begin{align}
  \label{eq:influx}
  F_{\mathrm{in}} & =N_f \int_{\mathbb{R}^d}(\bs{v}_2 - \bs{v}_s)   p_{\bs{r}_2,\bs{v}_2}(\bs{r}_s,\bs{v}_2) \; \mathrm{d} \bs{v}_2 \nonumber \\
                  & = N_f  \sum_{i = 1}^{m}   (\bs{v}^*_{2i} - \bs{v}_s) p_{\bs{r}2,i} (\bs{r}_s)
\end{align}
Then the impact rate to the spacecraft is
\begin{equation}
  \label{eq:impact_rate}
  \dot{\eta} (t) = A_c F_{\mathrm{in}}
\end{equation}
where $A_c$ is the cross-section area, and $\dot{\eta}$ means the number of impacts per unit time.

At last, we can evaluate the probability of collision between the debris cloud and the spacecraft over a given time interval:
\begin{equation}
  \label{eq:pc}
  P_c = 1 - e ^{-\eta}
\end{equation}
where
\begin{equation}
  \label{eq:eta}
  \eta =\int_{0}^{t} \dot{\eta} (t) \; \mathrm{d} t
\end{equation}
represents the cumulative number of impacts.

\subsection{Numerical Results}

The evolved distribution of clouds is then used to calculate the impact risk to spacecraft in GEO. 
The risks associated with fragments of different sizes are calculated separately, based on the number and velocity distribution detailed in Section \ref{sec:num_initial}.

\begin{figure}[htbp]
  \centering
  \includegraphics[width=1.0\textwidth]{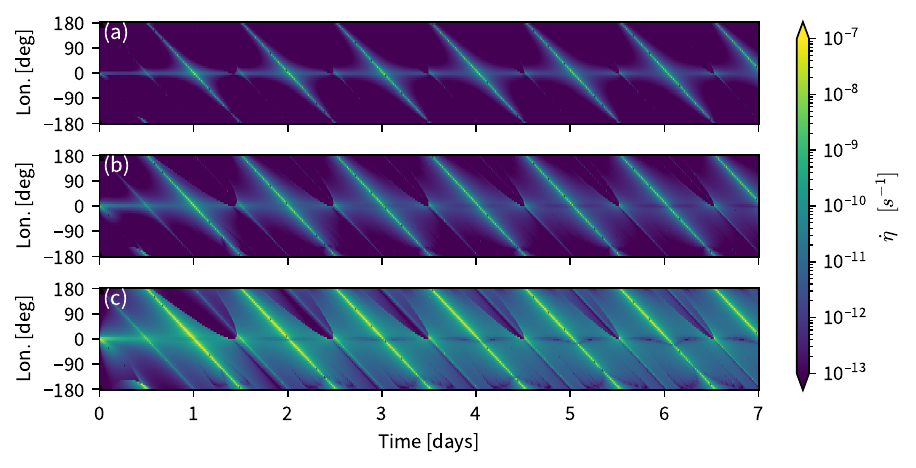}
  \caption{The Impact Rate of a Spacecraft in GEO}
  \label{fig:doteta}
\end{figure}

Figure \ref{fig:doteta} shows the impact rate to GEO spacecraft located at different longitude over 7 days. 
Panels (a) - (c) correspond size intervals of 5 cm to 1 m, 1 cm to 5 cm, and 1 mm to 1 cm, respectively.
As expected, the impact rate decreases for larger fragments due to their lower abundance (see Table \ref{tab:size_number}).
The impact rate is significantly higher near the fragmentation point and declines with increasing longitude. 
Despite this, the trend of the impact rate over time remains consistent.
The high impact rates are observed every second orbital period.
Spacecraft positioned east of the fragmentation point experience impacts sooner, as they are located ahead of the flight path of most fragments.

\begin{figure}[htbp]
  \centering
  \includegraphics[width=1.0\textwidth]{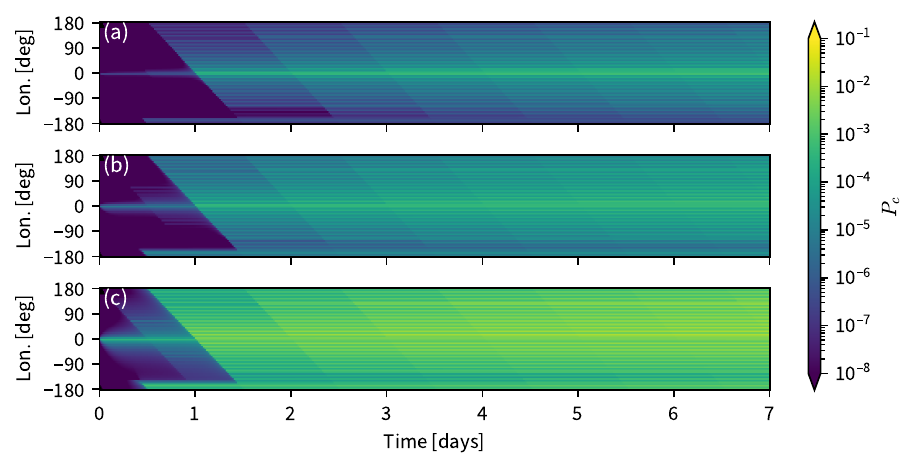}
  \caption{The Cumulative Collision Probability of a Spacecraft in GEO}
  \label{fig:pc_lon}
\end{figure}

Figure \ref{fig:pc_lon} shows the cumulative collision probability of spacecraft at different longitudes. 
Within 1.5 days, the probability of a collision with millimeter-sized fragments increases to $10^{-3}$ at most locations, while the probability of a collision with fragments of 5 cm or larger is approximately $10^{-6}$. 
Subsequently, the collision probability increases slowly because the impact rate is very low relative to $P_c$.

\begin{figure}[htbp]
  \centering
  \includegraphics[width=1.0\textwidth]{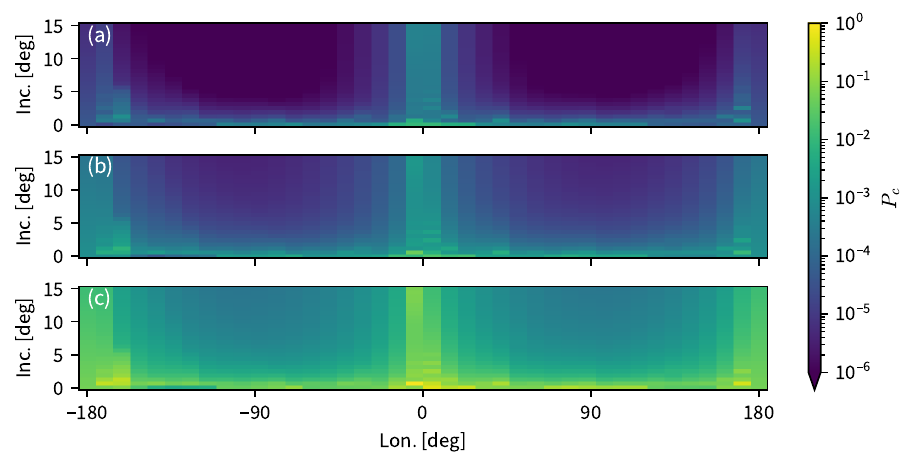}
  \caption{The Cumulative Collision Probability of a Spacecraft in GEO with Inclination}
  \label{fig:pc_inclon}
\end{figure}

Figure \ref{fig:pc_inclon} shows the cumulative collision probability for spacecraft across various longitudes and inclinations.
Only inclinations below 15 degrees are considered, as the majority of GEO objects fall within this range. 
The probability of collision decreases significantly for spacecraft at higher inclinations compared to those at lower inclinations. 
This difference arises because the fragments are primarily ejected from two parent orbits: those from the SC are concentrated in the equatorial plane, while those from the RB are mainly distributed along the 28.5-degree inclination plane (see Figure \ref{fig:pryz}).
However, exceptions occur at 0 and 180 degrees longitude, where the collision probability remains high even at higher inclinations. 
This can be attributed to the high density of fragments near the fragmentation point and along the diametrically opposite line.

\section{Conclusions}\label{sec:conclusions}
The short-term risk of debris clouds resulting from collision events in geostationary orbit was investigated.
The continuous distribution function of ejection velocities was integrated using the NASA Standard Breakup Model, and the initial distribution of the resulting two clouds was presented in a single probability density function.
A strong agreement was observed between this continuous density function and the discrete samples generated by an validated implementation of the breakup model. 
The three-dimensional distribution of the clouds exhibited a layered structure, with a high-density linear region opposite the fragmentation point.
The risk to geostationary spacecraft at various longitudes and inclinations was also evaluated. 
Numerical results indicated that objects east of the fragmentation point would encounter debris impacts sooner. 
The probability of collision with millimeter-sized fragments may increase to 1 \% within 36 hours, while the probability of collision with fragments of 5 cm or larger will approximate $10^{-5}$.
These results demonstrate the vulnerability of the densely populated geostationary region to space traffic accidents.

\section*{Declaration of competing interest}
The authors declare that they have no known competing financial interests or personal relationships that could have appeared to influence the work reported in this paper.

\section*{Acknowledgements} 
This work was supported by the National Natural Science Foundation of China under Grant Nos. 12303082, 12432017, and 12402048. 
Peng Shu also acknowledges support from the China Postdoctoral Science Foundation (CPSF) under Grant No. 2024M75349 and the Postdoctoral Fellowship Program of CPSF under Grant No. GZC20232975.

\appendix
\section{Probability Density and Number Density}
\label{sec:pd&nd}

A probability density function $p_{\bs{x}}$ is a non-negative Lebesgue-integrable function which satisfies the normalization condition:
\begin{equation}
    \label{eq:probability_density}
    \int_{\mathbb{R}^d} p_{\bs{x}}(\bs{x})\,\mathrm{d} \bs{x} =1
\end{equation}
where $\bs{x} \in \mathbb{R}^d$  is a $d$-dimensional random variable. The probability of finding $\bs{x}$ within a volume $V$ is then
\begin{equation}
    P(\bs{x} \in V)  = \int_{V} p_{\bs{x}}(\bs{x})\,\mathrm{d} \bs{x}
\end{equation}
If $\bs{x}$ is the position of a fragment and $V$ is a volume in the space, then $P(\bs{x} \in V)$ is the probability of finding a fragment in $V$.

Number density indicates the number of particles per unit volume in the vicinity of a specific location. If $n_{\bs{x}}$ is the number density of fragment clouds, then the number of fragments appearing in a volume $V$ is
\begin{equation}
    N(\bs{x} \in V)  = \int_{V} n_{\bs{x}}(\bs{x})\,\mathrm{d} \bs{x}
\end{equation}
where $n_{\bs{x}}$ is the number density, satisfying
\begin{equation}
    \int_{\mathbb{R}^d} n_{\bs{x}}(\bs{x})\,\mathrm{d} \bs{x} = N_f
\end{equation}
and $N_f$ is the total number of fragments.

\bibliographystyle{elsarticle-num} 
\bibliography{reference-origin}

\begin{thebibliography}{10}
\expandafter\ifx\csname url\endcsname\relax
  \def\url#1{\texttt{#1}}\fi
\expandafter\ifx\csname urlprefix\endcsname\relax\def\urlprefix{URL }\fi
\expandafter\ifx\csname href\endcsname\relax
  \def\href#1#2{#2} \def\path#1{#1}\fi

\bibitem{esasspacedebrisoffice2018Classification}
E.~S.~D. Office, Classification of geosynchronous objects, Technical {{Note}}
  GEN-DB-LOG-00211-OPS-GR, European Space Operations Centre, Darmstadt, Germany
  (May 2018).

\bibitem{mei2022Hybrid}
H.~Mei, Hybrid removal of end-of-life geostationary satellites using solar
  radiation pressure and impulsive thrusts, Ph.D. thesis (2022).

\bibitem{jewett2024Intelsats}
R.~Jewett, Intelsat's {{IS-33e Satellite}} is a `{{Total Loss}}',
  https://www.satellitetoday.com/connectivity/2024/10/21/intelsats-is-33e-satellite-is-a-total-loss/
  (Oct. 2024).

\bibitem{anz-meador2022History}
P.~D. {Anz-Meador}, J.~Opiela, J.-C. Liou, History of {{On-orbit Satellite
  Fragmentations}}, 16th {{Edition}}, Tech. Rep. NASA/TP-20220019160, NASA
  Orbital Debris Program Office, Houston, Texas (2022).

\bibitem{nafi2020Practical}
A.~M. Nafi, Practical optical survey strategies for near geostationary orbital
  debris, Ph.D. thesis (2020).

\bibitem{blake2021DebrisWatch}
J.~A. Blake, P.~Chote, D.~Pollacco, W.~Feline, G.~Privett, A.~Ash, S.~Eves,
  A.~Greenwood, N.~Harwood, T.~R. Marsh, D.~Veras, C.~Watson, {{DebrisWatch
  I}}: {{A}} survey of faint geosynchronous debris, Advances in Space Research
  67~(1) (2021) 360--370.
\newblock \href {https://doi.org/10.1016/j.asr.2020.08.008}
  {\path{doi:10.1016/j.asr.2020.08.008}}.

\bibitem{mcknight2019Space}
D.~McKnight, T.~Maclay, Space {{Environment Management}}: {{A Common Sense
  Framework}} for {{Controlling Orbital Debris Risk}}, in: Proceedings of the
  {{Advanced Maui Optical}} and {{Space Surveillance Technologies Conference}},
  The Maui Economic Development Board, Maui, Hawaii, 2019-09-17/2019-09-20, pp.
  1--8.

\bibitem{oltrogge2018Comprehensive}
D.~L. Oltrogge, S.~Alfano, C.~Law, A.~Cacioni, T.~S. Kelso, A comprehensive
  assessment of collision likelihood in {{Geosynchronous Earth Orbit}}, Acta
  Astronautica 147 (2018) 316--345.
\newblock \href {https://doi.org/10.1016/j.actaastro.2018.03.017}
  {\path{doi:10.1016/j.actaastro.2018.03.017}}.

\bibitem{programs2011Limiting}
C.~f. t. A. o. N. O.~D. Programs, N.~R. Council, Limiting {{Future Collision
  Risk}} to {{Spacecraft}}: {{An Assessment}} of {{NASA}}'s {{Meteoroid}} and
  {{Orbital Debris Programs}}, National Academies Press, Washington, 2011.

\bibitem{lawrence2022Case}
A.~Lawrence, M.~L. Rawls, M.~Jah, A.~Boley, F.~Di~Vruno, S.~Garrington,
  M.~Kramer, S.~Lawler, J.~Lowenthal, J.~McDowell, M.~McCaughrean, The case for
  space environmentalism, Nature Astronomy 6~(4) (2022) 428--435.
\newblock \href {https://doi.org/10.1038/s41550-022-01655-6}
  {\path{doi:10.1038/s41550-022-01655-6}}.

\bibitem{2018IADC}
{{IADC}} protection manual, Tech. Rep. IADC-04-03, Interagency Space Debris
  Coordination Committee (2018).

\bibitem{matney2019NASA}
M.~Matney, A.~Manis, P.~{Anz-Meador}, D.~Gates, J.~Seago, A.~Vavrin, Y.-L. Xu,
  The {{NASA Orbital Debris Engineering Model}} 3.1: {{Development}},
  {{Verification}}, and {{Validation}}, in: International {{Orbital Debris
  Conference}} ({{IOC}}), Sugar Land, TX, 2019.

\bibitem{krisko2014New}
P.~H. Krisko, The {{New NASA Orbital Debris Engineering Model ORDEM}} 3.0, in:
  {{AIAA}}/{{AAS Astrodynamics Specialist Conference}}, AIAA 2014-4227, San
  Diego, CA, 2014-08-04/2014-08-07.
\newblock \href {https://doi.org/10.2514/6.2014-4227}
  {\path{doi:10.2514/6.2014-4227}}.

\bibitem{horstmann2020Enhancement}
A.~Horstmann, Enhancement of {{S}}/{{C}} fragmentation and environment
  evolution models, Tech. Rep. DD-0045, Technische Universit{\"a}t of
  Braunschweig, Braunschweig (Aug. 2020).

\bibitem{krisko2015ORDEM}
P.~H. Krisko, S.~Flegel, M.~J. Matney, D.~R. Jarkey, V.~Braun, {{ORDEM}} 3.0
  and {{MASTER-2009}} modeled debris population comparison, Acta Astronautica
  113 (2015) 204--211.
\newblock \href {https://doi.org/10.1016/j.actaastro.2015.03.024}
  {\path{doi:10.1016/j.actaastro.2015.03.024}}.

\bibitem{liu2024Space}
Y.~Liu, R.~Chi, B.~Pang, {\relax HU}.~Diqi, W.~Cao, D.~Wang, Space debris
  environment engineering model 2019: {{Algorithms}} improvement and comparison
  with {{ORDEM}} 3.1 and {{MASTER-8}}, Chinese Journal of Aeronautics 37~(5)
  (2024) 392--409.
\newblock \href {https://doi.org/10.1016/j.cja.2023.12.004}
  {\path{doi:10.1016/j.cja.2023.12.004}}.

\bibitem{krisko2011Proper}
P.~H. Krisko, Proper implementation of the 1998 {{NASA}} breakup model, Orbital
  Debris Quarterly News 15~(4) (2011) 4--5.

\bibitem{mains2022IMPACT}
D.~L. Mains, M.~E. Sorge, The {{IMPACT}} satellite fragmentation model, Acta
  Astronautica 195 (2022) 547--555.
\newblock \href {https://doi.org/10.1016/j.actaastro.2022.03.030}
  {\path{doi:10.1016/j.actaastro.2022.03.030}}.

\bibitem{yao2024Generation}
T.~Yao, Z.~Yang, Y.~Luo, S.~Lan, L.~Ren, Generation of initial debris cloud
  distributions for breakup events based on {{CARDC-SBM}}, Acta Astronautica
  219 (2024) 580--591.
\newblock \href {https://doi.org/10.1016/j.actaastro.2024.03.060}
  {\path{doi:10.1016/j.actaastro.2024.03.060}}.

\bibitem{jang2024Simulating}
D.~Jang, R.~Linares, Simulating the {{Evolution}} of {{Lethal Non-Trackable
  Population}} and its {{Effect}} on {{LEO Sustainability}} (Aug. 2024).
\newblock \href {https://doi.org/10.48550/arXiv.2408.15025}
  {\path{doi:10.48550/arXiv.2408.15025}}.

\bibitem{liou2006Risks}
J.-C. Liou, N.~L. Johnson, Risks in {{Space}} from {{Orbiting Debris}}, Science
  311~(5759) (2006) 340--341.
\newblock \href {https://doi.org/10.1126/science.1121337}
  {\path{doi:10.1126/science.1121337}}.

\bibitem{au2001Estimation}
S.-K. Au, J.~L. Beck, Estimation of small failure probabilities in high
  dimensions by subset simulation, Probabilistic Engineering Mechanics 16~(4)
  (2001) 263--277.
\newblock \href {https://doi.org/10.1016/S0266-8920(01)00019-4}
  {\path{doi:10.1016/S0266-8920(01)00019-4}}.

\bibitem{mcinnes1993Analytical}
C.~R. McInnes, An analytical model for the catastrophic production of orbital
  debris, ESA Journal 17~(4) (1993) 293--305.

\bibitem{letizia2015Analytical}
F.~Letizia, C.~Colombo, H.~G. Lewis, Analytical {{Model}} for the
  {{Propagation}} of {{Small-Debris-Object Clouds After Fragmentations}},
  Journal of Guidance, Control, and Dynamics 38~(8) (2015) 1478--1491.
\newblock \href {https://doi.org/10.2514/1.G000695}
  {\path{doi:10.2514/1.G000695}}.

\bibitem{letizia2016Collision}
F.~Letizia, C.~Colombo, H.~G. Lewis, Collision {{Probability Due}} to {{Space
  Debris Clouds Through}} a {{Continuum Approach}}, Journal of Guidance,
  Control, and Dynamics 39~(10) (2016) 2240--2249.
\newblock \href {https://doi.org/10.2514/1.G001382}
  {\path{doi:10.2514/1.G001382}}.

\bibitem{frey2021Transformation}
S.~Frey, C.~Colombo, Transformation of {{Satellite Breakup Distribution}} for
  {{Probabilistic Orbital Collision Hazard Analysis}}, Journal of Guidance,
  Control, and Dynamics 44~(1) (2021) 88--105.
\newblock \href {https://doi.org/10.2514/1.G004939}
  {\path{doi:10.2514/1.G004939}}.

\bibitem{giudici2024Densitybaseda}
L.~Giudici, C.~Colombo, A.~Horstmann, F.~Letizia, S.~Lemmens, Density-based
  evolutionary model of the space debris environment in low-{{Earth}} orbit,
  Acta Astronautica 219 (2024) 115--127.
\newblock \href {https://doi.org/10.1016/j.actaastro.2024.03.008}
  {\path{doi:10.1016/j.actaastro.2024.03.008}}.

\bibitem{wen2024Modeling}
C.~Wen, Z.~Jin, C.~Peng, D.~Qiao, Modeling {{Medium-Term Debris Cloud}} of
  {{Satellite Breakup}} via {{Probabilistic Method}}, Journal of Guidance,
  Control, and Dynamics 47~(8) (2024) 1602--1619.
\newblock \href {https://doi.org/10.2514/1.G008062}
  {\path{doi:10.2514/1.G008062}}.

\bibitem{healy2016Structure}
L.~M. Healy, S.~Kindl, E.~Rolfe, C.~Binz, Structure and evolution of a debris
  cloud in the early phases, in: Advances in the {{Astronautical Sciences}},
  Vol. 158, Univelt, Inc., San Diego, CA, 2016, pp. 2715--2743.
\newblock \href {https://doi.org/10.5281/zenodo.1406548}
  {\path{doi:10.5281/zenodo.1406548}}.

\bibitem{healy2020Orbital}
L.~M. Healy, C.~R. Binz, S.~Kindl, Orbital {{Dynamic Admittance}} and {{Earth
  Shadow}}, The Journal of the Astronautical Sciences 67~(2) (2020) 427--457.
\newblock \href {https://doi.org/10.1007/s40295-018-00144-1}
  {\path{doi:10.1007/s40295-018-00144-1}}.

\bibitem{vallado2017Fragmentation}
D.~A. Vallado, D.~L. Oltrogge, Fragmentation {{Event Debris Field Evolution
  Using 3D Volumetric Risk Assessment}}, in: 7th {{European Conference}} on
  {{Space Debris}}, ESA Space Debris Office, Darmstadt, Germany,
  2017-04-18/2017-04-21.

\bibitem{oltrogge2017Application}
D.~L. Oltrogge, D.~A. Vallado, Application of new debris risk evolution and
  dispersal ({{DREAD}}) tool to characterize post-fragmentation risk, in:
  {{AAS}}/{{AIAA Astrodynamics Specialist Conference}}, Vol. 162, 2017, pp.
  463--482.

\bibitem{shu2022Collision}
P.~Shu, Z.~Yang, Y.-Z. Luo, Z.-J. Sun, Collision {{Probability}} of {{Debris
  Clouds Based}} on {{Higher-Order Boundary Value Problems}}, Journal of
  Guidance, Control, and Dynamics 45~(8) (2022) 1512--1522.
\newblock \href {https://doi.org/10.2514/1.G006356}
  {\path{doi:10.2514/1.G006356}}.

\bibitem{shu2023Impact}
P.~Shu, Z.~Yang, Y.-Z. Luo, Impact {{Risk}} of a {{Debris Cloud}} to
  {{Spacecraft}}, Journal of Guidance, Control, and Dynamics 46~(5) (2023)
  989--997.
\newblock \href {https://doi.org/10.2514/1.G007056}
  {\path{doi:10.2514/1.G007056}}.

\bibitem{parigini2024ShortTerm}
C.~Parigini, R.~Algethamie, R.~Armellin, Short-{{Term Collision Probability
  Caused}} by {{Debris Cloud}}, Journal of Guidance, Control, and Dynamics
  47~(5) (2024) 874--886.
\newblock \href {https://doi.org/10.2514/1.G007687}
  {\path{doi:10.2514/1.G007687}}.

\bibitem{francesconi2019CST}
A.~Francesconi, C.~Giacomuzzo, L.~Olivieri, G.~Sarego, M.~Duzzi, F.~Feltrin,
  A.~Valmorbida, K.~D. Bunte, M.~Deshmukh, E.~Farahvashi, J.~Pervez, M.~Zaake,
  T.~Cardone, D.~{de Wilde}, {{CST}}: {{A}} new semi-empirical tool for
  simulating spacecraft collisions in orbit, Acta Astronautica 160 (2019)
  195--205.
\newblock \href {https://doi.org/10.1016/j.actaastro.2019.04.035}
  {\path{doi:10.1016/j.actaastro.2019.04.035}}.

\bibitem{olivieri2024Numerical}
L.~Olivieri, C.~Giacomuzzo, A.~Francesconi, Numerical simulation of {{COSMOS}}
  2499 fragmentation, CEAS Space Journal 16~(6) (2024) 659--665.
\newblock \href {https://doi.org/10.1007/s12567-024-00545-z}
  {\path{doi:10.1007/s12567-024-00545-z}}.

\bibitem{schuhmacher2021Efficient}
J.~Schuhmacher, Efficient {{Implementation}} and {{Evaluation}} of the {{NASA
  Breakup Model}} in modern {{C}}++, Bachelor's {{Thesis}}, Technical
  University of Munich (2021).

\end{thebibliography}

\end{document}